\documentclass[twocolumn,aps,prb,showpacs,amsmath,amssymb,10pt]{revtex4-1}
\usepackage{amsmath}
\usepackage{cases}
\usepackage{amssymb}
\usepackage[dvips]{graphicx}
\usepackage{color}
\usepackage{multirow}
\newcommand{\vect}[1]{\boldsymbol{#1}}

\begin{document}

\title{Force and heat current formulas for many-body potentials in molecular dynamics simulation with applications to thermal conductivity calculations}
\author{Zheyong Fan$^{1,2}$}
\email{brucenju@gmail.com}
\author{Luiz Felipe C. Pereira$^{3}$}
\email{pereira@dfte.ufrn.br}
\author{Hui-Qiong Wang$^{4}$}
\author{Jin-Cheng Zheng$^{5}$}
\author{Davide Donadio$^{6,7,8}$}
\author{Ari Harju$^{2}$}
\affiliation{$^{1}$School of Mathematics and Physics, Bohai University, Jinzhou, China}
\affiliation{$^{2}$COMP Centre of Excellence, Department of Applied Physics, Aalto University, Helsinki, Finland}
\affiliation{$^{3}$Departamento de F\'isica Te\'orica e Experimental, Universidade Federal do Rio Grande do Norte, Natal, RN, 59078-900, Brazil}
\affiliation{$^{4}$Key Laboratory of Semiconductors and Applications of Fujian Province, Department of Physics, Xiamen University, Xiamen, P.R. China 361005}
\affiliation{$^{5}$Fujian Provincial Key Laboratory of Mathematical Modeling and High-Performance Scientific Computation, Department of Physics, Xiamen University, Xiamen, P.R. China 361005}
\affiliation{$^{6}$Max Planck Institut f\"ur Polymerforschung, Ackermannweg 10, D-55128 Mainz, Germany}
\affiliation{$^{7}$Donostia International Physics Center, Paseo Manuel de Lardizabal, 4, 20018 Donostia-San Sebastian, Spain}
\affiliation{$^{8}$IKERBASQUE, Basque Foundation for Science, E-48011 Bilbao, Spain}

\date{\today}

\begin{abstract}
We derive expressions of interatomic force and heat current for many-body potentials such as the Tersoff, the Brenner, and the Stillinger-Weber potential used extensively in molecular dynamics simulations of covalently bonded materials. Although these potentials have a many-body nature, a pairwise force expression that follows Newton's third law can be found without referring to any partition of the potential. Based on this force formula, a stress applicable for periodic systems can be unambiguously defined. The force formula can then be used to derive the heat current formulas using a natural potential partitioning. Our heat current formulation is found to be equivalent to most of the seemingly different heat current formulas used in the literature, but to deviate from the stress-based formula derived from two-body potential. We validate our formulation numerically on various systems descried by the Tersoff potential, namely three-dimensional silicon and diamond, two-dimensional graphene, and quasi-one-dimensional carbon nanotube. The effects of cell size and time used in the simulation are examined.
\end{abstract}

\maketitle

\section{Introduction}

Molecular dynamics (MD) simulation has been used extensively to study thermal transport properties of materials. There are mainly two methods for computing lattice thermal conductivity in the level of classical MD simulations: the direct method \cite{jund1999,muller1997} [also called the nonequilibrium MD (NEMD) method] based on the Fourier's law and the Green-Kubo \cite{green1954,kubo1957,mcquarrie2000} method (also called the equilibrium MD method) based on the Green-Kubo formula. Cross-checking of these two methods has also been the subject of several works. \cite{schelling2002,sellan2010,he2012} In the direct method, the thermal conductivity is usually computed by measuring the steady-state temperature gradient at a fixed external heat current, analogous to the experimental situation. In contrast, in the Green-Kubo method, the thermal conductivity is computed by integrating the heat current autocorrelation function (HCACF) using the Green-Kubo formula. While the heat current in the direct method is created by scaling the velocities in the source and sink regions of the simulated system, which does not depend on the underline interatomic potential, the heat current in the Green-Kubo method is the summation of the microscopic heat currents of the individual atoms in the simulated system, which generally depends on the specific interatomic potential used.

For a two-body potential, where a pairwise force can be directly defined, the heat current expression used in the Green-Kubo formula is well established. It is currently implemented in Large-scale Atomic/Molecular Massively Parallel Simulator (LAMMPS) \cite{plimption1995} in terms of the per-atom stress and works well for systems described by two-body potentials such as Lennard-Jones argon. However, it is not widely recognized that the heat current expression based on the per-atom stress is only applicable to two-body potentials, and is not guaranteed to produce correct results for systems described by a many-body potential, such as the widely used Tersoff potential, \cite{tersoff1989} Brenner potential, \cite{brenner1990} and Stillinger-Weber potential. \cite{stillinger1985} In the literature, there have been quite a few formulations \cite{li1998,che2000,berber2000,dong2001,chen2006} of the heat current for the Tersoff/Brenner potential, which seem to be inequivalent to each other. \cite{cuellar2010,khadem2013}

In this work, we present detailed derivations of the heat current expressions for these many-body potentials. We show that many of the seemingly different formulations of the heat current are equivalent, except for some marginal differences resulting from a different decomposition of the total potential into site (per-atom) potentials. Our derivation
is facilitated by establishing the existence of a pairwise force respecting Newton's third law, which is not widely recognized so far. Based on the pairwise force, a well defined expression for the virial tensor can also be obtained. By comparing with finite-difference calculations, we validate the proposed pairwise force expression unambiguously. The derived expression is equivalent to other alternatives which do not respect Newton¡¯s third law explicitly, but it has an advantage of allowing for an efficient implementation on graphics processing units (GPUs), which attains a speedup factor of two orders of magnitude (compared to our optimized serial CPU code) for large simulation cell sizes.

Using the efficient GPU code, we perform a comprehensive validation of our formulations by calculating lattice thermal conductivities of various kinds of material described by the Tersoff potential, including three-dimensional (3D) silicon and diamond, two-dimensional (2D) graphene, and quasi-one-dimensional (Q1D) carbon nanotube (CNT). For each material, we examine the convergence of the calculated thermal conductivity with respect to the total simulation time, the correlation time, and the finite-size effects, before comparing our results with previous ones. Last, we present explicit numerical evidence that the stress-based heat current expression is inequivalent to our formulation for the Tersoff potential.

\section{Theory}

\subsection{Green-Kubo method for thermal conductivity calculations}

The Green-Kubo formula for the running thermal conductivity (RTC) tensor $\kappa_{\mu\nu}(t)$ ($\mu, \nu = x, y, z$) at a given correlation time $t$ can be expressed as \cite{green1954,kubo1957,mcquarrie2000}
\begin{equation}
\label{equation:Green-Kubo}
\kappa_{\mu\nu}(t) = \frac{1}{k_B T^2 V} \int_0^{t} dt' C_{\mu\nu} (t'),
\end{equation}
where $k_B$ is Boltzmann's constant, $T$ is the absolute temperature, and $V$ is the volume of the simulation cell. The HCACF $C_{\mu\nu}(t)$ is defined as
\begin{equation}
\label{equation:HCACF}
 C_{\mu\nu}(t) = \langle J_{\mu}(t=0) J_{\nu}(t) \rangle,
\end{equation}
the ensemble average of the product of two heat currents separated by $t$. In the MD simulation, the ensemble average is substituted by a time average. The simulation time required for achieving high statistical accuracy of the computed thermal conductivity in the Green-Kubo method is usually quite challenging, as we show later. The Green-Kubo method is capable of calculating the full conductivity tensor, but the following cases are sufficient to verify our formulations: (1) isotropic 3D systems, such as diamond, where we define the conductivity scalar as $(\kappa_{xx} + \kappa_{yy} + \kappa_{zz})/3$, (2) isotropic 2D systems, such as graphene, where we define the in-plane conductivity as $(\kappa_{xx} + \kappa_{yy})/2$, and (3) Q1D systems, such as CNT, where only the conductivity along the tube is needed. Periodic boundary conditions are needed in all the transport directions. In the following, we use $\vect{J}$ to represent the heat current vector with components $J_x$, $J_y$, and $J_z$.

\subsection{General expression of the heat current}

The heat current used in Eq.~(\ref{equation:HCACF}) is defined as the time derivative of the sum of the moments of the site energies
\begin{equation}
E_i = \frac{1}{2} m_i \vect{v}_i^2 + U_i
\end{equation}
of the particles in the system \cite{mcquarrie2000}:
\begin{equation}
  \vect{J} \equiv \frac{d}{d t} \sum_i \vect{r}_i E_i
  = \sum_i \vect{v}_i E_i + \sum_i \vect{r}_i \frac{d}{d t} E_i.
\end{equation}
Here $m_i$, $\vect{v}_i$, and $U_i$ are the mass, velocity, and potential energy of particle $i$, respectively. Conventionally, one defines a kinetic part
\begin{equation}
\vect{J}_{\textmd{kin}} =  \sum_i \vect{v}_i E_i
\end{equation}
and a potential part
\begin{equation}
\vect{J}_{\textmd{pot}} =  \sum_i \vect{r}_i \frac{d}{d t} E_i
\end{equation}
and write the total heat current as a sum of them:
\begin{equation}
\vect{J} =  \vect{J}_{\textmd{kin}} + \vect{J}_{\textmd{pot}}.
\end{equation}
The kinetic term  $\vect{J}_{\textmd{kin}}$ needs no further derivation, apart from a possible issue of defining $U_i$ for a many-body potential, and the potential term  $\vect{J}_{\textmd{pot}}$ can be written as
\begin{equation}
  \vect{J}_{\textmd{pot}} = \sum_i \vect{r}_i (\vect{F}_i \cdot \vect{v}_i)
  + \sum_i \vect{r}_i \frac{d U_{i}}{d t},
  \label{equation:j_pot}
\end{equation}
where the kinetic energy theorem, $\frac{d}{dt}\left(\frac{1}{2}m_i\vect{v}_i^2\right)=\vect{F}_i \cdot \vect{v}_i$, $\vect{F}_i$ being the total force on particle $i$, has been used. The kinetic term is also called the convective term, and is mostly important for gases. For Lennard-Jones liquid, Vogelsang \textit{et al}.\cite{vogelsang1987} showed that the thermal conductivity is mainly contributed by the partial HCACF involving the potential-potential term. For solids, the kinetic term barely contributes and can be simply discarded. Note that the kinetic and potential terms defined here correspond to the potential and kinetic terms, respectively, used in the Einstein formalism studied by Kinaci \textit{et al}., \cite{kinaci2012} who also found that the convective term (the potential term in the Einstein formalism) does not contribute to the thermal conductivity for solids. We thus focus on the potential part [Eq.~(\ref{equation:j_pot})] in the following discussions.

\subsection{Heat current for two-body potentials}

Before discussing many-body potentials, let us first examine the case of two-body potentials. For these, the total potential energy of the system can be written as
\begin{equation}
  U =  \frac{1}{2} \sum_{i}  \sum_{j \neq i} U_{ij} ,
  \label{equation:potential_pair}
\end{equation}
where the pair potential between particles $i$ and $j$, $U_{ij} = U_{ji} = U_{ij}(r_{ij})$, only depends on the distance $r_{ij}$ between the particles. The factor of $1/2$ in the above equation compensates the double-counting of the pair potentials; one can equally omit it by requiring $j>i$ (or $j<i$). The derived forces are purely
pairwise and Newton's third law is apparently valid:
\begin{equation}
  \vect{F}_i = \sum_{j \neq i} \vect{F}_{ij},
\end{equation}
\begin{equation}
\label{equation:force_pair}
  \vect{F}_{ij}
  = \frac{\partial U_{ij}}{\partial \vect{r}_{ij}}
  = - \vect{F}_{ji},
\end{equation}
where $\vect{F}_{ij}$ is the force on particle $i$ due to particle $j$ and the convention, \cite{note_convention}
\begin{equation}
 \label{equation:r_ij}
 \vect{r}_{ij} \equiv \vect{r}_{j} - \vect{r}_{i},
\end{equation}
for the relative position between two particles is adopted. If periodic boundary conditions are applied in a given direction, the minimum image convention is used to all the relative positions in that direction. Using the above notations, the first term on the right hand side (RHS) of Eq.~(\ref{equation:j_pot}) can be written as
\begin{equation}
\sum_i \vect{r}_i (\vect{F}_i \cdot \vect{v}_i)
= \sum_i \sum_{j \neq i} \vect{r}_i (\vect{F}_{ij} \cdot \vect{v}_i).
\end{equation}
To make further derivation for the second term  on the RHS of Eq.~(\ref{equation:j_pot}), one has to make a choice for the site potential $U_i$. A natural choice is $U_i = \frac{1}{2}\sum_{j\neq i} U_{ij}$, but for two-body potentials, it does not matter much how to define the site potential. For example, the above choice is equivalent to $U_i = \frac{1}{4}\sum_{j\neq i} (U_{ij}+U_{ji})$ because $U_{ij} = U_{ji}$. Therefore, the second term on the RHS of Eq.~(\ref{equation:j_pot}) can be written as
\begin{equation}
\sum_i \vect{r}_i \frac{d U_{i}}{d t}
= \frac{1}{2}\sum_i \sum_{j \neq i}
\vect{r}_i [\vect{F}_{ij} \cdot (\vect{v}_j-\vect{v}_i)].
\end{equation}
Using the above two expressions, we can write the potential term of the heat current as
\begin{equation}
\label{equation:j_pot_pair_1}
  \vect{J}^{\textmd{pair}}_{\textmd{pot}}
  = \frac{1}{2}\sum_i \sum_{j \neq i}
    \vect{r}_{i} [\vect{F}_{ij} \cdot (\vect{v}_i + \vect{v}_j)].
\end{equation}
In numerical calculations, the absolute positions, $\vect{r}_i$, will cause problems for systems with periodic boundary conditions. Fortunately, one can circumvent the difficulty by using the Newton's third law Eq.~(\ref{equation:force_pair}), from which we have
\begin{equation}
  \label{equation:j_pot_pair_2}
  \vect{J}^{\textmd{pair}}_{\textmd{pot}}
  = - \frac{1}{4}\sum_i \sum_{j \neq i}
    \vect{r}_{ij} [\vect{F}_{ij} \cdot (\vect{v}_i + \vect{v}_j)],
\end{equation}
where only the relative positions, $\vect{r}_{ij}$, are involved. This expression is also equivalent to a less symmetric form:
\begin{equation}
\label{equation:j_pot_pair_3}
  \vect{J}^{\textmd{pair}}_{\textmd{pot}}
  = - \frac{1}{2}\sum_i \sum_{j \neq i}
    \vect{r}_{ij} [\vect{F}_{ij} \cdot \vect{v}_i].
\end{equation}
The potential part of the heat current is also intimately related to the virial part of the stress tensor. To see this, we first note that the virial $\textbf{W}$ can be written as a summation of individual terms,
\begin{equation}
 \textbf{W} = \sum_i \textbf{W}_{i},
\end{equation}
where the per-atom virial $\textbf{W}_i$ for a periodic system reads
\begin{equation}
 \label{equation:per_atom_virial}
 \textbf{W}_i
 = -\frac{1}{2} \sum_{j \neq i}\vect{r}_{ij} \otimes \vect{F}_{ij}.
\end{equation}
Therefore, the potential part of the heat current can be expressed in terms of the per-atom virial as
\begin{equation}
\label{equation:j_pot_pair_stress}
  \vect{J}^{\textmd{stress}}_{\textmd{pot}} = \sum_{i} \textbf{W}_{i} \cdot \vect{v}_i.
\end{equation}
The current implementation of the Green-Kubo formula for thermal conductivity in LAMMPS adopts this stress-based formula. However, as we show later, it does not apply to many-body potentials.

\subsection{Force expressions for Tersoff potential}

We now move on to many-body potentials, first focusing on the Tersoff potential. The total potential energy for a system described by the Tersoff potential can also be written in the form of Eq. (\ref{equation:potential_pair}), with the ``pair potential" $U_{ij}$ taking the form of \cite{tersoff1989}
\begin{equation}
\label{equation:Tersoff_U_ij}
U_{ij} = f_C(r_{ij}) \left[ f_R(r_{ij}) - b_{ij} f_A(r_{ij}) \right],
\end{equation}
\begin{equation}
b_{ij} = \left(1 + \beta^n \zeta^n_{ij}\right)^{-\frac{1}{2n}},
\end{equation}
\begin{equation}
\zeta_{ij} = \sum_{k\neq i, j}f_C(r_{ik}) g_{ijk},
\end{equation}
\begin{equation}
g_{ijk} = 1 + \frac{c^2}{d^2} - \frac{c^2}{d^2+(h-\cos\theta_{ijk})^2}.
\end{equation}
Here, $\beta$, $n$, $c$, $d$, and $h$ are parameters and $\theta_{ijk}$ is the angle formed by $\vect{r}_{ij}$ and $\vect{r}_{ik}$, which means that
\begin{equation}
\label{equation:cos_theta}
\cos\theta_{ijk} = \cos\theta_{ikj} = \frac{\vect{r}_{ij} \cdot \vect{r}_{ik}}{r_{ij} r_{ik}}.
\end{equation}
For simplicity, the dependence of the parameters on the particle type is omitted in the above equations. Detailed rules for determining the parameters in systems with two kinds of atom can be found in Ref. [\onlinecite{tersoff1989}]. While the functions $f_{C}$, $f_{R}$, and $f_{A}$ only depend on $r_{ij}$, the bond-order function $b_{ij}$ also depends on the positions $\vect{r}_k$ of the neighbor particles of $i$ and $j$ and thus generally, $U_{ij} \neq U_{ji}$, which is a manifestation of the many-body nature of the Tersoff potential. However, we notice that $b_{ij}$, hence $U_{ij}$, is only a function of the position difference vectors originating
from particle $i$ (In the equation below, $k=j$ is allowed.):
\begin{equation}
 \label{equation:crucial_property}
 U_{ij} = U_{ij}
 \left(\{\vect{r}_{ik}\}_{k\neq i}\right).
\end{equation}
This property will play a crucial role in the following derivations.

We now start to derive the force expressions for the Tersoff potential. We begin with the definition:
\begin{equation}
 \label{equation:before_expansion}
 \vect{F}_i \equiv - \frac{\partial U}{\partial \vect{r}_i}
 \equiv -\frac{1}{2} \sum_j\sum_{k\neq j} \frac{\partial U_{jk}}{\partial \vect{r}_i}.
\end{equation}
We can expand it as
\begin{equation}
\label{equation:after_expansion}
\vect{F}_i = - \frac{1}{2} \left( \sum_{k\neq i} \frac{\partial U_{ik}}{\partial \vect{r}_i}
                                + \sum_{j\neq i} \frac{\partial U_{ji}}{\partial \vect{r}_i}
                                + \sum_{j\neq i}\sum_{k\neq j,i} \frac{\partial U_{jk}}{\partial \vect{r}_i}  \right).
\end{equation}
The first, second, and third terms on the RHS of Eq.~(\ref{equation:after_expansion}) correspond to the parts with
$j=i$, $k=i$, and $j, k \neq i$ in Eq.~(\ref{equation:before_expansion}), respectively. Then, using Eq.~(\ref{equation:crucial_property}), we have
\begin{align}
\vect{F}_i
=& - \frac{1}{2} \left( \sum_{k\neq i}\sum_{j\neq i} \frac{\partial U_{ik}}{\partial \vect{r}_{ij}}
     \frac{\partial \vect{r}_{ij}}{\partial \vect{r}_{i}}
     + \sum_{j\neq i}\sum_{k\neq j} \frac{\partial U_{ji}}{\partial \vect{r}_{jk}}
     \frac{\partial \vect{r}_{jk}}{\partial \vect{r}_{i}} \right) \nonumber \\
& - \frac{1}{2} \sum_{j\neq i}\sum_{k\neq j,i} \sum_{m\neq j}\frac{\partial U_{jk}}{\partial \vect{r}_{jm} }
     \frac{\partial \vect{r}_{jm}}{\partial \vect{r}_{i}} \nonumber\\
=&   \frac{1}{2} \left( \sum_{k\neq i}\sum_{j\neq i} \frac{\partial U_{ik}}{\partial \vect{r}_{ij}}
     + \sum_{j\neq i} \frac{\partial U_{ji}}{\partial \vect{r}_{ij}}
     + \sum_{j\neq i}\sum_{k\neq j,i} \frac{\partial U_{jk}}{\partial \vect{r}_{ij} } \right).
\end{align}
Since
\begin{equation}
\sum_{k\neq i}\sum_{j\neq i} \frac{\partial U_{ik}}{\partial \vect{r}_{ij}}
= \sum_{k\neq i, j}\sum_{j\neq i} \frac{\partial U_{ik}}{\partial \vect{r}_{ij}}
+ \sum_{j\neq i} \frac{\partial U_{ij}}{\partial \vect{r}_{ij}},
\end{equation}
we have
\begin{equation}
\label{equation:F_i_complicated}
\vect{F}_i = \frac{1}{2}\sum_{j\neq i} \frac{\partial}{\partial \vect{r}_{ij}}
\left( U_{ij} + U_{ji} + \sum_{k\neq i, j} \left(U_{ik} + U_{jk}\right) \right).
\end{equation}
From this, a pairwise force between two particles can also be defined for the many-body Tersoff potential:
\begin{equation}
\label{equation:F_ij}
\vect{F}^{\text{Tersoff}}_{ij} \equiv \frac{1}{2}\frac{\partial}{\partial \vect{r}_{ij}}
\left( U_{ij} + U_{ji} + \sum_{k\neq i, j} \left(U_{ik} + U_{jk}\right) \right).
\end{equation}
The total force can be expressed as a sum of the pairwise forces
\begin{equation}
\vect{F}_i = \sum_{j \neq i} \vect{F}^{\text{Tersoff}}_{ij},
\end{equation}
and Newton's third law
\begin{equation}
\vect{F}^{\text{Tersoff}}_{ij} = - \vect{F}^{\text{Tersoff}}_{ji}
\end{equation}
still holds.

In the above derivations, we have not assumed any form of the site potential $U_i$. The definition of $U_i$ for a many-body potential amounts to a decomposition of the total potential into site potentials. While such a decomposition is not needed for the derivation of the forces, it is needed for deriving the heat current, which involves time-derivative of the site potential [cf. Eq.~(\ref{equation:j_pot})]. A natural choice for the decomposition is
\begin{equation}
\label{equation:U_i}
 U = \sum_{i} U_{i} \qquad \text{with}\qquad U_i \equiv \frac{1}{2} \sum_{j \neq i} U_{ij}.
\end{equation}
There is no clear physical intuition favoring this decomposition over others [cf. Eq.~(\ref{equation:U_i_Li})], but we find that Eq.~(\ref{equation:U_i}) is a very reasonable definition. To show this, we notice that the site potential defined by Eq.~(\ref{equation:U_i}) is also only a function of the relative positions originating from particle $i$:
\begin{equation}
\label{equation:crucial_property_again}
U_{i} = U_{i}
\left( \{\vect{r}_{ij}\}_{j\neq i}\right).
\end{equation}
Using this property, the total force on particle $i$ can be derived as
\begin{align}
\vect{F}_i
\label{equation:F_i_simplified}
&\equiv - \frac{\partial U}{\partial \vect{r}_i}
\equiv -\sum_j \frac{\partial U_j}{\partial \vect{r}_i}
= -\sum_{j\neq i} \left(\frac{\partial U_j}{\partial \vect{r}_i}\right)
  -\frac{\partial U_i}{\partial \vect{r}_i} \nonumber \\
&= -\sum_{j\neq i} \left( \sum_{k\neq j}\frac{\partial U_j}{\partial \vect{r}_{jk}}
    \frac{\partial \vect{r}_{jk}}{\partial \vect{r}_i}
   +\frac{\partial U_i}{\partial \vect{r}_{ij}}
    \frac{\partial \vect{r}_{ij}}{\partial \vect{r}_i} \right) \nonumber\\
&=   \sum_{j \neq i}
    \left(
    \frac{\partial U_i}{\partial \vect{r}_{ij}} -
    \frac{\partial U_j}{\partial \vect{r}_{ji}}
    \right),
\end{align}
which is equivalent to Eq.~(\ref{equation:F_i_complicated}), and the pairwise force is simplified to be
\begin{equation}
\label{equation:F_ij_simplified}
\vect{F}^{\text{Tersoff}}_{ij} =
    \left(
    \frac{\partial U_i}{\partial \vect{r}_{ij}} -
    \frac{\partial U_j}{\partial \vect{r}_{ji}}
    \right).
\end{equation}

One can check that Eq.~(\ref{equation:F_ij_simplified}) reduces to Eq.~(\ref{equation:force_pair}) in the case of two-body interaction. We also point out that our force expressions for the Tersoff potential are only seemingly different from other alternatives. There should be no ambiguity for the calculation of the total force on a given particle. However, different formulations may lead to different computer implementations. A crucial advantage of our formulation is that the total forces for individual particles can be calculated independently, which is desirable for massively parallel implementation. The numerical calculations presented in this work were performed by a molecular dynamics code implemented on GPUs using the thread-scheme in Ref. [\onlinecite{fan2013}]. However, a detailed presentation of the GPU-implementation of the Tersoff potential is beyond the scope this paper, which will be presented elsewhere.

Another advantage of our formulation is that the per-atom virial for the Tersoff potential takes the same form as for two-body potential:
\begin{equation}
\label{equation:per_atom_virial_tersoff}
\textbf{W}^{\text{Tersoff}}_i = -\frac{1}{2} \sum_{j \neq i}\vect{r}_{ij}
\otimes \vect{F}^{\text{Tersoff}}_{ij}
\end{equation}
which is unambiguously defined for periodic systems. \cite{louwerse2006} This might not be exactly equivalent to what has been implemented in LAMMPS, where the per-atom virial is calculated as
\begin{align}
\label{equation:per_atom_virial_lammps}
\textbf{W}_i =
& -\frac{1}{2} \sum_{j \neq i}\vect{r}_{ij} \otimes \vect{F}^{(2)}_{ij} \nonumber \\
& -\frac{1}{3} \sum_{j \neq i} \sum_{k \neq i, j}
\left(
\vect{r}_{ij} \otimes \vect{F}^{(3)}_{ij} + \vect{r}_{ik} \otimes \vect{F}^{(3)}_{ik}
\right).
\end{align}
Here, $\vect{F}^{(2)}_{ij}$, $\vect{F}^{(3)}_{ij}$, and $\vect{F}^{(3)}_{ik}$ represent the force components on particle $i$ associated with the two-body part due to particle $j$, the three-body part due to particle $j$, and the three-body part due to particle $k$, respectively. Although Eq.~(\ref{equation:per_atom_virial_lammps}) and Eq.~(\ref{equation:per_atom_virial_tersoff}) may result in the same total virial tensor for the Tersoff potential, they may not be equivalent when used to compute the heat current and lattice thermal conductivity. We will present numerical results to compare them.

\subsection{Heat current for the Tersoff potential}

We now derive the heat current expressions for the Tersoff potential, using the potential decomposition given by Eq.~(\ref{equation:U_i}). Using Eq.~(\ref{equation:F_i_simplified}), the first term on the RHS of Eq.~(\ref{equation:j_pot}) can be written as
\begin{equation}
\sum_i \vect{r}_i (\vect{F}_i \cdot \vect{v}_i)
= \sum_i \sum_{j \neq i} \vect{r}_i
\left(\frac{\partial U_i}{\partial \vect{r}_{ij}}-
\frac{\partial U_j}{\partial \vect{r}_{ji}}\right)
\cdot \vect{v}_i.
\end{equation}
Using Eq.~(\ref{equation:crucial_property_again}), the second term on the RHS of Eq.~(\ref{equation:j_pot}) can be written as
\begin{equation}
\sum_i \vect{r}_i \frac{d U_{i}}{d t}
= \sum_i \sum_{j \neq i}
\vect{r}_i
\frac{\partial U_i}{\partial \vect{r}_{ij}}
\cdot (\vect{v}_j-\vect{v}_i).
\end{equation}
From these two expressions, we get the following formula for the potential part of the heat current for the Tersoff potential:
\begin{equation}
\vect{J}^{\textmd{Tersoff}}_{\textmd{pot}} = \sum_i \sum_{j \neq i} \vect{r}_{i}
\left(
      \frac{\partial U_i}{\partial \vect{r}_{ij}} \cdot \vect{v}_j
    - \frac{\partial U_j}{\partial \vect{r}_{ji}} \cdot \vect{v}_i
\right).
\label{equation:j_pot_tersoff_1}
\end{equation}
Again, one can get rid of the absolute positions $\vect{r}_i$ by rewriting the above formula as:
\begin{equation}
\vect{J}^{\textmd{Tersoff}}_{\textmd{pot}} = - \frac{1}{2}\sum_i \sum_{j \neq i} \vect{r}_{ij}
\left(
      \frac{\partial U_i}{\partial \vect{r}_{ij}} \cdot \vect{v}_j
    - \frac{\partial U_j}{\partial \vect{r}_{ji}} \cdot \vect{v}_i
\right).
\label{equation:j_pot_tersoff_2}
\end{equation}
A less symmetric form can also be readily obtained:
\begin{equation}
\vect{J}^{\textmd{Tersoff}}_{\textmd{pot}} = - \sum_i \sum_{j \neq i} \vect{r}_{ij}
\left(
      \frac{\partial U_i}{\partial \vect{r}_{ij}} \cdot \vect{v}_j
\right),
\label{equation:j_pot_tersoff_3}
\end{equation}
or equivalently,
\begin{equation}
\vect{J}^{\textmd{Tersoff}}_{\textmd{pot}} = \sum_i \sum_{j \neq i} \vect{r}_{ij}
\left(
      \frac{\partial U_j}{\partial \vect{r}_{ji}} \cdot \vect{v}_i
\right).
\label{equation:j_pot_tersoff_4}
\end{equation}

Therefore, the potential part of the heat current for the Tersoff potential is not equivalent to the stress-based formula given by Eq.~(\ref{equation:j_pot_pair_stress}). One can check that, in the case of two-body interactions, the heat current expressions in Eqs. (\ref{equation:j_pot_tersoff_1}-\ref{equation:j_pot_tersoff_4})
for the Tersoff potential reduce to those for the two-body potential in Eqs.~(\ref{equation:j_pot_pair_1}-\ref{equation:j_pot_pair_3}).

Apart from the velocities $\vect{v}_i$ and relative positions $\vect{r}_{ij}$, the only nontrivial terms in the force and heat current expressions are $\frac{\partial U_i}{\partial \vect{r}_{ij}}$ and $\frac{\partial U_j}{\partial \vect{r}_{ji}}$, the latter being able to be obtained from the former by an exchange of $i$ and $j$. An explicit expression for the former is presented in appendix \ref{appendix:explicit_expression}.

In appendix \ref{appendix:hardy}, we show that Eq.~(\ref{equation:j_pot_tersoff_4}) is equivalent to the one derived by Hardy \cite{hardy1963} at the quantum level for general many-body interactions. In the following, we refer to Eq.~(\ref{equation:j_pot_tersoff_4}) as the Hardy formula and Eq.~(\ref{equation:j_pot_pair_stress}) as the stress formula.

There has been some confusion about the seemingly different heat current expressions for the Tersoff potential in the literature. Guajardo-Cu\'ellar \textit{et al.} \cite{cuellar2010} and Khadem \textit{et al.} \cite{khadem2013} compared several expressions \cite{li1998,che2000,dong2001,chen2006,cuellar2010,hardy1963} in the literature. From their results, it seems as if all of these expressions were inequivalent. In appendix \ref{appendix:hardy}, we show that many of them are equivalent to the Hardy formula.

\subsection{Generalization to other many-body potentials}

Besides the Tersoff potential, the Brenner potential \cite{brenner1990} and the Stillinger-Weber (SW) potential \cite{stillinger1985} are also widely used in the study of covalently bonded systems. Here, we first show that the derivations for the Tersoff potential can be generalized to these potentials and then summarize our results for a general many-body potential.

The generalization to the Brenner potential is straightforward. The pair potential $U_{ij}$ for this takes the same form as that for the Tersoff potential [Eq.~(\ref{equation:Tersoff_U_ij})]. The bond-order function $b_{ij}$, hence $U_{ij}$, is only a function of the position difference vectors originating from particle $i$, although the explicit form of $b_{ij}$ in the Brenner potential is more complicated. This is the only property we used to derive the pairwise force expression [Eq.~(\ref{equation:F_ij})] for the Tersoff potential. Therefore, the same pairwise force expression also applies to the Brenner potential. Using the same potential partition as for the Tersoff potential, $U_i=\frac{1}{2}\sum_{j \neq i} U_{ij}$, we can arrive at a simplified pairwise force expression [Eq.~(\ref{equation:F_ij_simplified})] and the Hardy formula [Eq.~(\ref{equation:j_pot_tersoff_4})] of heat current, as in the case of the Tersoff potential.

We next consider the SW potential. The total potential energy consists of a two-body part and a three-body part, the latter being given as: \cite{stillinger1985}
\begin{equation}
U^{(3)} = \sum_{i}\sum_{j> i}\sum_{k>j}(h_{ijk}+h_{jki}+h_{kij}),
\end{equation}
where
\begin{equation}
h_{ijk}=\lambda \exp\left[ \frac{\gamma}{r_{ij}-a} + \frac{\gamma}{r_{ik}-a} \right]
         \left(\cos \theta_{ijk} + \frac{1}{3} \right)^2.
\end{equation}
Here, $\lambda$, $\gamma$, and $a$ are parameters and $\cos \theta_{ijk}$ is defined as in Eq.~(\ref{equation:cos_theta}). Similar definitions apply to $h_{jki}$ and $h_{kij}$. It is clear that $h_{ijk}$ is symmetric in the last two indices: $h_{ijk} = h_{ikj}$. Using this property, we can reexpress the three-body part of the total potential as
\begin{equation}
U^{(3)} = \frac{1}{6}\sum_{i}\sum_{j\neq i}\sum_{k\neq i,j}(h_{ijk}+h_{jki}+h_{kij}),
\end{equation}
which can be further simplified as
\begin{equation}
U^{(3)} = \frac{1}{2}\sum_{i}\sum_{j\neq i}\sum_{k\neq i,j}h_{ijk}.
\end{equation}
Without referring to any potential partition, but noticing that $h_{ijk}$ is only a function of the position difference vectors originating from particle $i$, one can derive a pairwise force expression for the three-body part:
\begin{equation}
\vect{F}^{(3)}_{i} = \sum_{j \neq i} \vect{F}^{(3)}_{ij},
\end{equation}
\begin{align}
\vect{F}^{(3)}_{ij} &=\frac{1}{2}
\left(
     \sum_{k\neq i}\sum_{m\neq i, k} \frac{\partial h_{ikm}}{\partial \vect{r}_{ij}}
    + \sum_{k\neq j}\sum_{m\neq j, k} \frac{\partial h_{jkm}}{\partial \vect{r}_{ij}}
\right) \nonumber \\
&= -\vect{F}^{(3)}_{ji}.
\end{align}
With a definition of the site potential,
\begin{equation}
\label{equation:U_i_SW}
 U^{(3)}_i \equiv \frac{1}{2} \sum_{j\neq i}\sum_{k\neq i,j}h_{ijk} \quad
     \text{with} \quad U^{(3)} = \sum_{i} U^{(3)}_{i},
\end{equation}
the above pairwise force expression can be simplified to
\begin{equation}
\vect{F}^{(3)}_{ij} =
    \frac{\partial U^{(3)}_i}{\partial \vect{r}_{ij}} -
    \frac{\partial U^{(3)}_j}{\partial \vect{r}_{ji}}.
\end{equation}
This is formally the same as that for the Tersoff potential, the only difference being the form of the site potential. Adopting the above potential decomposition, and noticing that $U_i$ is only a function of the position difference vectors originating from particle $i$, one can confirm that the potential part of the heat current also takes the form of the Hardy formula:
\begin{equation}
\vect{J}^{(3)}_{\textmd{pot}} = \sum_i \sum_{j \neq i} \vect{r}_{ij}
\left(
      \frac{\partial U^{(3)}_j}{\partial \vect{r}_{ji}} \cdot \vect{v}_i
\right).
\end{equation}

In fact, the pairwise force formula and the Hardy formula of heat current apply to any many-body potential, because the crucial property we have used in the above derivations, i.e., the pair potential $U_{ij}$ (or the site potential $U_{i}$) is only a function of the set of vectors $\{\vect{r}_{ij}\}_{j \neq i}$, is satisfied by any empirical potential: any other position difference vector can be expressed as the difference of two vectors in this set. In other words, the vectors $\{\vect{r}_{ij}\}_{j \neq i}$ form a complete set of independent arguments for any pair or site potential associated with particle $i$. We can summarize our formulations as follows. For a general classical many-body potential,
\begin{equation}
U=\sum_i U_i \left( \{\vect{r}_{ij}\}_{j \neq i} \right),
\end{equation}
there exists a pairwise force between two particles $i$ and $j$,
\begin{equation}
\label{equation:pair_force_many_body}
\vect{F}_{ij} = -\vect{F}_{ji} =
    \frac{\partial U_i}{\partial \vect{r}_{ij}} -
    \frac{\partial U_j}{\partial \vect{r}_{ji}},
\end{equation}
a well defined virial tensor for periodic systems,
\begin{equation}
\label{equation:virial_many_body}
 \textbf{W}
 = -\frac{1}{2} \sum_i \sum_{j \neq i}\vect{r}_{ij} \otimes \vect{F}_{ij},
\end{equation}
and a well defined potential part of the heat current for periodic systems,
\begin{equation}
\label{equation:heat_many_body}
\vect{J}_{\textmd{pot}} = \sum_i \sum_{j \neq i} \vect{r}_{ij}
\left(
      \frac{\partial U_j}{\partial \vect{r}_{ji}} \cdot \vect{v}_i
\right).
\end{equation}

The existence of a pairwise force for classical many-body potentials, albeit not surprising according to the principles of classical mechanics, has not been widely recognized in the community. Without an explicit expression for the pairwise force, much effort has been devoted to constructing general expressions for the virial tensor in periodic systems. \cite{louwerse2006,thompson2009} Our formulations are thus not only useful for thermal conductivity calculations based on the Green-Kubo formula, but can also find application in the study of properties related to the stress tensor.

\section{Validating the pairwise force expression}
One of the interesting results from the previous section is that the interatomic forces for empirical many-body potentials are in fact totally pairwise. Here, we take the Tersoff potential an example and present numerical evidence for the correctness of the pairwise force by comparing the forces calculated using Eq.~(\ref{equation:F_i_simplified})
and those using the finite-difference formula,
\begin{equation}
\label{equation:force_finite_difference}
 F_{ix} = \frac{
                U(\cdots, \vect{r}_i - \Delta x \vect{e_x}, \cdots) -
                U(\cdots, \vect{r}_i + \Delta x \vect{e_x}, \cdots)
               } {2 \Delta x}.
\end{equation}
Here, $F_{ix}$ is the $x$-component of the total force on particle $i$ when the system is in a specific configuration and $\Delta x \vect{e_x}$ is a small displacement vector of particle $i$ along the $x$-direction
from its original position $\vect{r}_i$. One can similarly consider the other directions, but the results for all the directions are quite similar and we only present the results for a single direction for simplicity. We have checked that the forces calculated by the finite-difference method do not change over a wide range of $\Delta x$.

\begin{figure}
\begin{center}
  \includegraphics[width=\columnwidth]{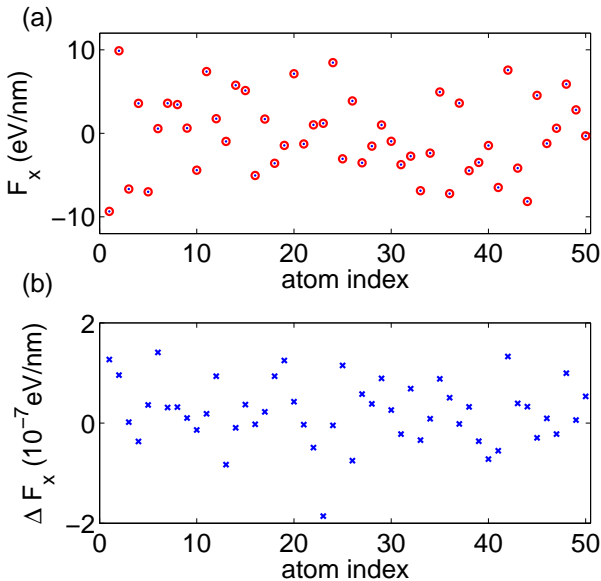}
  \caption{(Color online) (a) Total forces in the $x$-direction on individual
           carbon atoms in a configuration generated
           by randomly shifting the positions of all the atoms from the perfect
           graphene structure by a small amount. The small solid dots and the larger open circles
           represent the results by using Eqs.~(\ref{equation:F_i_simplified})
           and (\ref{equation:force_finite_difference}), respectively.
           (b) Force differences between those obtained by using
           Eqs.~(\ref{equation:F_i_simplified})
           and (\ref{equation:force_finite_difference}) as shown in (a).
           Note that the testing system has more than 50 atoms, but only
           the data for the first 50 atoms are shown for clarity.
           Results for the $y$- and $z$-directions are similar.
  }
\label{figure:force_comparision}
\end{center}
\end{figure}

The results for the comparison are shown in Fig.~\ref{figure:force_comparision}. The system corresponds to a graphene sheet perturbed from the perfect honeycomb structure by randomly shifting the positions of all the atoms by a small amount. One can see that the forces on each particle calculated by the pairwise force expression Eq.~(\ref{equation:F_i_simplified}) and the finite-difference expression Eq.~(\ref{equation:force_finite_difference}) are practically the same, with the relative errors being as small as about $10^{-8}$. This comparison thus confirms the correctness of the pairwise force for the Tersoff potential unambiguously.

\section{Applications on thermal conductivity calculations}

With the force expression validated, we are now in a position to apply the heat current formulations to study lattice thermal conductivities of various kinds of material. To be specific, we present results obtained by using the Tersoff potential, which has been applied extensively in the study of thermal transport properties of silicon, diamond, graphene, and CNT. The Tersoff parameters used for diamond and silicon are taken from Ref. [\onlinecite{tersoff1989}] and those for graphene and CNT are the optimized ones obtained by Lindsay and Broido. \cite{lindsay2010} To be specific, we only consider isotopically pure $^{12}$C and $^{28}$Si in our simulations, although our method is not limited to this case. When calculating the thermal conductivity of graphene and CNT, one has to specify the effective thickness of the graphene sheet. We have chosen it to be 0.335 nm. We use cubic simulation cells for silicon and diamond and roughly square-shaped simulation cells for graphene. The time step of integration in the MD simulations is chosen to be 1 fs for most of the simulated systems, but for smaller carbon systems, we found that smaller time steps are desirable. The evolution time in the equilibration stage (canonical ensemble, where temperature is controlled) of the MD simulation lasts one to several nanoseconds, depending on the simulations cell size. The heat current data are recorded every 10 steps in the production stage (microcanonical ensemble, where temperature is not controlled).  We only consider systems with zero external pressure and the lattice constants for silicon at 500 K and diamond at 300 K are determined to be 0.544 nm and 0.357 nm. For grahene and CNT at 300 K, the average carbon-carbon distance is determined to be 0.144 nm.

\subsection{Silicon}

\begin{figure}[htb]
\begin{center}
\includegraphics[width=\columnwidth]{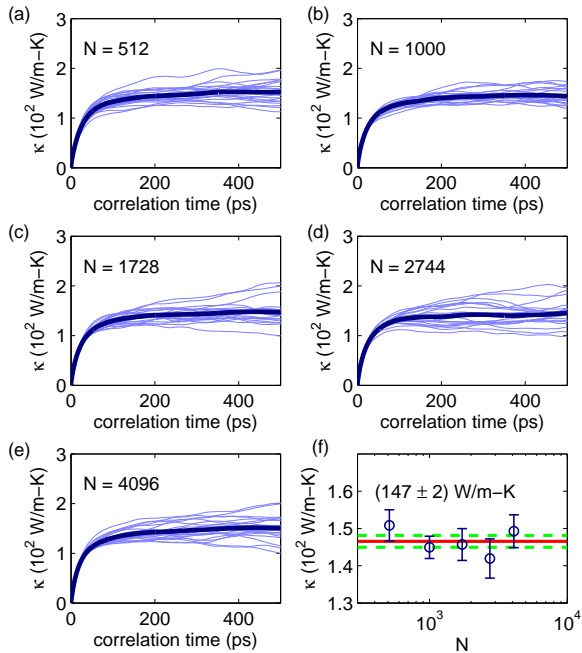}
\caption{(Color online) (a-e) Running thermal conductivities as a function of correlation time for silicon with different simulation cell sizes at 500 K. The thinner (and lighter) and the thicker (and darker) lines represent the results of independent simulations with different initial velocities and the ensemble average over the independent simulations, respectively. (f) Thermal conductivity as a function of the simulation cell size $N$. Markers with error bars represent the average values and the corresponding standard errors for a given $N$. The solid line indicates the average ($147$ W/m-K) over the 5 simulation cell sizes and the dashed lines indicate the corresponding standard error ($\pm 2$ W/m-K).}
\label{figure:silicon}
\end{center}
\end{figure}

We start presenting our results by considering silicon. Figs.~\ref{figure:silicon}(a-e) show the RTCs [given by Eq.~(\ref{equation:Green-Kubo})] for silicon at 500 K with different simulation cell sizes $N$. For a given $N$, there are large variations between the independent simulations associated with different sets of initial velocities in the MD simulations. Despite the variations, a well converged RTC can be obtained by averaging over sufficiently many independent simulations, along with estimations of an average value and the corresponding error estimate for the converged thermal conductivity. In this work, we determine them in the following steps (for a given $N$):
\begin{enumerate}
\item Determine (by visual inspection) a range of correlation time $[t_1, t_2]$ where the averaged RTC has converged well.
\item Calculate the average values of the RTCs for the independent simulations over the range of correlation time determined in the last step.
\item Take the mean value and standard error (standard deviation divided by $\sqrt{M}$, where $M$ is the number of independent simulations) of the average values obtained in the last step as the average value and error estimate, which are represented by an open circle and the corresponding error bar in Fig.~\ref{figure:silicon}(f) for a given $N$.
\end{enumerate}
To determine $[t_1, t_2]$, we have to ensure that the averaged RTC is sufficiently smooth. The smoothness can be enhanced by increasing either the simulation time $t_s$ of the individual simulations or the number of independent simulations $N_s$. More precisely, it is determined by the product $N_s t_s$. We found that a value of $N_s t_s=$ 200 ns is enough for silicon at 500 K. It can be seen that all the averaged RTCs in Figs.~\ref{figure:silicon}(a-e) are rather smooth and $[t_1, t_2]=$ [400 ps, 500 ps] is a fairly good choice for the converged time interval.

Before comparing our results with previous ones, we need to further check possible finite-size effects in the calculations. The Green-Kubo formula is, in principle, only meaningful for infinite systems, i.e., systems in the thermodynamic limit. However, in practice, one can only simulate systems with finite simulation cell sizes, with periodic boundary conditions applied along the directions which are thought to be infinite to alleviate the finite-size effects in those directions. One can then check if the results converge with increasing simulation cell size.

Figure~\ref{figure:silicon}(f) presents the converged thermal conductivities of silicon at 500 K obtained by using different simulation cell sizes: $N=$ 512, 1000, 1728, 2744, and 4096. It can be seen that they do not show a systematical decreasing or increasing trend with increasing $N$.

Due to the small finite-size effects, we can take the average values of thermal conductivity for different simulation cell sizes as independent simulation results and obtain an average value and the corresponding error estimate. In this way, we obtain the final result, $(147 \pm 2)$ W/m-K, which is in good agreement with that obtained by Howell, \cite{howell2012} $(155\pm4)$ W/m-K. Note that Howell used the direct method with the same Tersoff parameters. This comparison thus further confirmed the equivalence between the direct method and the Green-Kubo method, as has been shown by Schelling \textit{et al.} \cite{schelling2002} for SW silicon.

\subsection{Diamond}

\begin{figure}[htb]
\begin{center}
\includegraphics[width=\columnwidth]{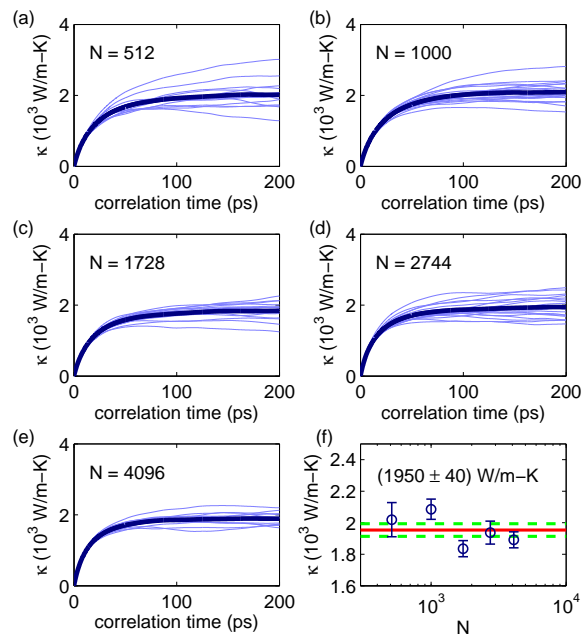}
\caption{(Color online) Same as Fig.~\ref{figure:silicon}, but for diamond at 300 K.}
\label{figure:diamond}
\end{center}
\end{figure}

We next consider diamond. The RTCs at 300 K with 5 simulation cell sizes, $N=$ 512, 1000, 1728, 2744, and 4096, are shown in Figs.~\ref{figure:diamond}(a-e) and the corresponding converged values are presented in Fig.~\ref{figure:diamond}(f). The averaged RTCs converge earlier than those for silicon. Here, it can be seen that the converged time interval can be chosen to be $[t_1, t_2] =$ [150 ps, 200 ps]. Due to the shorter correlation time required for converging, the total simulation time required for obtaining smooth curves of the RTC is shorter than that for silicon, being about $N_s t_s =$ 100 ns.

As in the case of silicon, there is no systematical decreasing or increasing trend with increasing $N$. Our calculated thermal conductivity averaged over the 5 simulation cell sizes is $(1950 \pm 40)$ W/m-K. Using the Brenner potential \cite{brenner1990} and the Green-Kubo method, Che \textit{et al.} \cite{che2000} obtained a converged value of about 1200 W/m-K for isotopically pure $^{12}$C diamond, which is about one third smaller than ours. This difference can be understood by noticing that the original Brenner potential is more anharmonic than the original Tersoff potential, as has also been noticed in the study of CNT and graphene. \cite{lindsay2010} Experimentally, the thermal conductivity of isotopically  pure $^{12}$C diamond at room temperature is about 3000 W/m-K, \cite{wei1993} larger than both of our results. The difference between theoretical and experimental results may result from an excessive anharmonicity of the empirical potentials.

\subsection{Graphene}

\begin{figure}[htb]
\begin{center}
\includegraphics[width=\columnwidth]{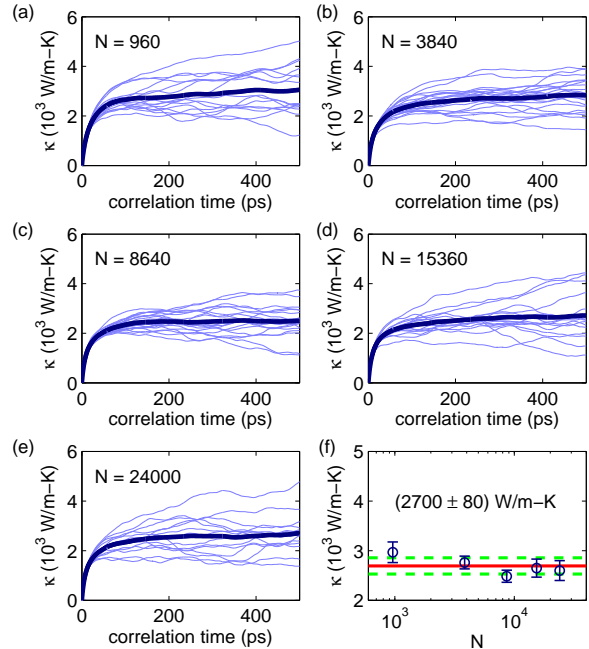}
\caption{(Color online) Same as Fig.~\ref{figure:silicon}, but for graphene at 300 K.}
\label{figure:graphene}
\end{center}
\end{figure}

The above results are for 3D bulk materials. We now turn to study low-dimensional materials, first considering 2D graphene. The RTCs at 300 K with 5 simulation cell sizes, $N=$ 960, 3840, 8640, 15360, and 24000, are shown in Figs.~\ref{figure:graphene}(a-e), with the corresponding converged values presented in Fig.~\ref{figure:graphene}(f). For each $N$, a total simulation time of $N_st_s =$ 500 ns is required to obtain an average RTC well converged in the time interval of $[t_1, t_2]=$ [250 ps, 500 ps].

As in the case of diamond and silicon, the thermal conductivity of graphene does not increase with increasing simulation cell size. In fact, the contrary is true when $N$ is smaller than $10^4$, as found by Pereira and Donadio. \cite{pereira2013prb} Similar results have also been obtained by Zhang \textit{et al.} \cite{zhang2011} for smaller $N$. The increasing of the simulation cell size has two opposite effects: (1) It allows more long-wavelength phonons, which can increase the thermal conductivity; (2) It also allows more phonon scattering, as suggested \cite{ladd1986} by Ladd \textit{et al}., which can decrease the thermal conductivity. In 2D graphene, more phonon scattering can be induced by the acoustic flexural modes with increasing out-of-plane deformation, which is positively correlated to the simulation cell size. \cite{fasolino2007} When the simulation cell size is relatively small, the second effect may dominate, resulting in a decreasing thermal conductivity with increasing simulation cell size. When the simulation cell size is relatively large, these two effects largely compensate each other, resulting in converged thermal conductivity with increasing simulation cell size.

The thermal conductivity of graphene at 300 K averaged over the 5 simulation cell sizes is $(2700 \pm 80)$ W/m-K. Using the optimized Brenner potential \cite{lindsay2010} and the Green-Kubo method, Zhang \textit{et al.} \cite{zhang2011, note_potential} reported a converged value of $(2900 \pm 93)$ W/m-K for graphene at 300 K, which is slightly larger than ours. This difference may be explained by the fact that they have used smaller simulation cell sizes, which, according to the discussion above, results in larger thermal conductivity for graphene. On the other hand, Haskins \textit{et al.} \cite{haskins2011} reported a value of 2600 W/m-K based on the Einstein formulation, \cite{kinaci2012} which is in good agreement with ours.

It is interesting to point out that our estimate of the thermal conductivity for graphene at room temperature is compatible with NEMD calculations (using the same Tersoff potential parameters) in Ref. [\onlinecite{xu2014}], which give $\kappa \approx 2300$ W/m-K with a simulation length of about 1.5 $\mu$m. If we take the consistency between the Green-Kubo method and the NEMD method as granted, this comparison indicates that the NEMD results have not been converged up to a simulation length of 1.5 $\mu$m. In fact, both the NEMD results and the experimental data \cite{xu2014} suggest a logarithmic length-dependence of thermal conductivity of graphene at the micrometer scale. On the other hand, whether the thermal conductivity is upper-limited or not in the infinite-size limit has been largely debated recently. \cite{xu2014,lindsay2014,fugallo2014,Barbarino2015} Our results provide evidence that the thermal conductivity of an extended (macroscopic) graphene sheet is finite, although at the micrometer scale $\kappa$ still depends on the length of the graphene patch.

\subsection{(10, 0)-carbon nanotube}

\begin{figure}[htb]
\begin{center}
\includegraphics[width=\columnwidth]{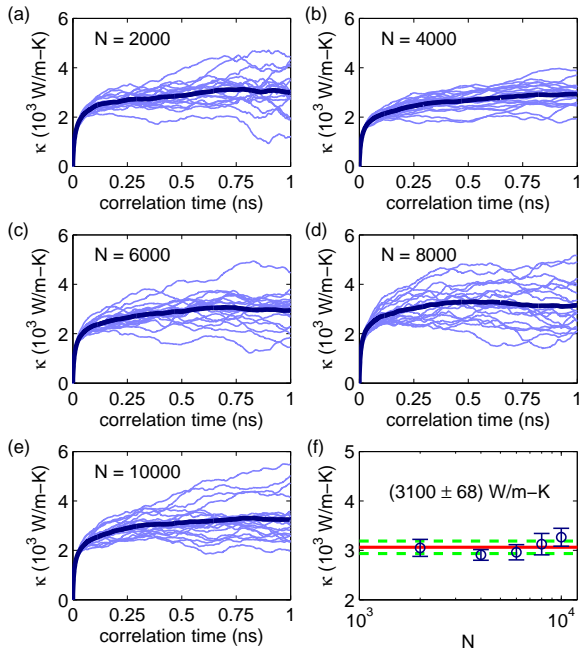}
\caption{(Color online) Same as Fig.~\ref{figure:silicon}, but for (10, 0)-CNT at 300 K.}
\label{figure:cnt}
\end{center}
\end{figure}

Last, we examine the longitudinal thermal conductivity of CNT. To be specific, we consider a (10, 0)-CNT, without a detailed study of the effects of chirality and radius.
The RTCs at 300 K with 5 simulation cell sizes, $N=$ 2000, 4000, 6000, 8000, and 10000, are shown in Figs.~\ref{figure:cnt}(a-e), with the corresponding converged values presented in Fig.~\ref{figure:cnt}(f). For each $N$, a total simulation time of $N_st_s =$ 1000 ns is required to obtain an average RTC almost converged in the time interval of $[t_1, t_2]=$ [500 ps, 1000 ps].

Compared with 2D graphene, the (10, 0)-CNT has even larger thermal conductivity: $(3100\pm68)$ W/m-K. This high value of thermal conductivity is mostly due to the long phonon wavelength (large phonon relaxation time) in Q1D CNTs, \cite{mahan2004} as indicated by the slow convergence of $\kappa$ with respect to $t$. While there were debates on the size convergence of $\kappa$ for CNTs, \cite{mingo2005,donadio2007,savin2009,lindsay2009,donadio2009} our results do not suggest a divergent $\kappa$ with respect to the simulation cell length. Previously, the thermal conductivity for (10, 0)-CNT was calculated to be $(1750 \pm 230)$ W/m-K in Ref. [\onlinecite{donadio2007}] (see also Ref. [\onlinecite{donadio2009}]) and $(1700 \pm 200)$ W/m-K in Ref. [\onlinecite{pereira2013njp}], which are both smaller than the value obtained in this work, but due to different reasons: Ref. [\onlinecite{donadio2007}] employed the original parameter set provided by Tersoff, \cite{tersoff1989}; Ref. [\onlinecite{pereira2013njp}] used the stress formula as implemented in LAMMPS, which also results in smaller values of $\kappa$ comparing with the Hardy formula, as we show below.

\subsection{Comparing the stress and the Hardy formula}

\begin{figure}[htb]
\begin{center}
\includegraphics[width=\columnwidth]{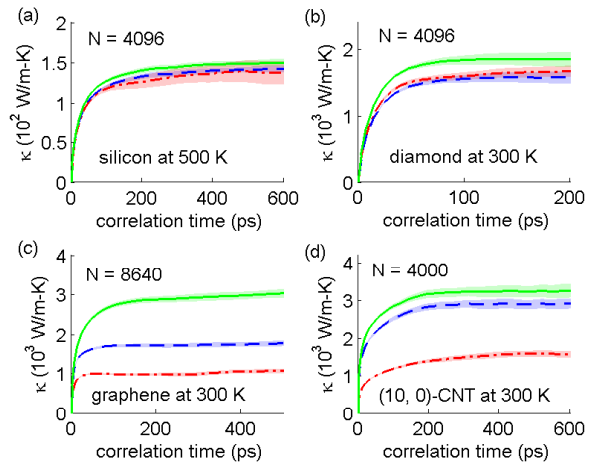}
\caption{(Color online) Running thermal conductivities $\kappa(t)$ as a function of correlation time for (a) silicon at 500 K, (b) diamond at 300 K, (c) graphene at 300 K, and (d) (10, 0)-CNT at 300 K obtained by using the Hardy formula (solid lines), the stress formula (dashed lines) and LAMMPS (dot-dashed lines). For each material, the line and the shaded area represent the averaged $\kappa(t)$ and the standard error calculated from an ensemble of 10 independent simulations. }
\label{figure:compare_stress_and_Hardy}
\end{center}
\end{figure}

Previously, we remarked that the stress formula [Eq. (\ref{equation:j_pot_pair_stress})] and the Hardy formula [Eq.~(\ref{equation:j_pot_tersoff_4})] are inequivalent for the Tersoff potential. Also, the per-atom virial as implemented in LAMMPS  [Eq.~(\ref{equation:per_atom_virial_lammps})] is not likely to be equivalent to ours [Eq.~(\ref{equation:per_atom_virial_tersoff})], which would result in different heat currents based on the stress formula. Here, we show the nonequivalence numerically.

Figure~\ref{figure:compare_stress_and_Hardy} shows the RTCs of (a) silicon at 500 K, (b) diamond at 300 K, (c) graphene at 300 K, and (d) (10,~0)-CNT at 300 K calculated using the Hardy formula, the stress formula in our formulation, and the stress formula as implemented in LAMMPS. We note the following observations based on Fig.~\ref{figure:compare_stress_and_Hardy}:

(1) For 3D diamond and silicon, all the three methods result in comparable results.

(2) For 2D graphene, the RTC in the converged regime ([250 ps, 500 ps]) obtained by the stress formula is about 1/2 of that by the Hardy formula and that by the LAMMPS implementation is about 1/3 of that by the Hardy formula. The LAMMPS results are consistent with previous ones. \cite{pereira2013prb,wang2014}

(3) For Q1D CNT, while the RTC in the converged regime ([400 ps, 600 ps]) obtained by the stress formula is comparable to  that by the Hardy formula, that by the LAMMPS implementation is about 1/2 of that by the Hardy formula. The LAMMPS results are also consistent with previous ones. \cite{pereira2013njp}

From these observations, we conclude that the stress formula is generally inequivalent to the Hardy formula, and the LAMMPS implementation of the stress formula is also inequivalent to our implementation based on the pairwise force. Although we are not clear about the reason why the differences between these formulations are more significant in low-dimensional materials (especially 2D graphene) than in 3D materials, our results can explain an extraordinary low value of thermal conductivity of graphene at 300 K, $(280 \pm 15)$ W/m-K, obtained by Mortazavi \textit{et al.} \cite{mortazavi2014} using LAMMPS and the (second-generation) Brenner potential. \cite{brenner2002} Apart from the higher anharmonicity of this empirical potential compared with the optimized Tersoff potential, this small thermal conductivity could be attributed to the use of the stress formula implemented in LAMMPS.

\section{Conclusions}
In summary, we formulated force, stress, and heat current expressions of many-body potentials in MD simulations. After deriving these expressions for the Tersoff potential in detail and briefly discussing their generalizations to the Brenner potential and the Stillinger-Weber potential, we reached a set of universal expressions [Eqs.~(\ref{equation:pair_force_many_body}-\ref{equation:heat_many_body})] which apply to general many-body potentials.

The pairwise force expression [Eq.~(\ref{equation:pair_force_many_body})], whose existence is guaranteed by the principles of classical mechanics, has not been widely recognized in the community so far. We presented explicit numerical evidence of its correctness by comparing with finite-difference calculations and demonstrated its importance in the construction of a well defined virial tensor [Eq.~(\ref{equation:virial_many_body})]. With a reasonable potential partition, we arrived at the Hardy formula [Eq.~(\ref{equation:heat_many_body})] for the potential part of microscopic heat current used in lattice thermal conductivity calculations based on the Green-Kubo formula. Many of the seemingly different formulations of the heat current in the literature were demonstrated to be equivalent to the Hardy formula.

We have implemented the formulations for the Tersoff potential on GPUs and obtained orders of magnitude speedup compared to our serial CPU implementation. While the details of the GPU-implementation is beyond the scope of this paper, we have applied it to calculate systematically the lattice thermal conductivities of various kinds of material, including 3D silicon and diamond, 2D graphene, and Q1D CNT, with emphasis on the effects of the simulation time and simulation cell size. We demonstrated the correctness of our formulations by comparing our results with previous ones. Last, we provided explicit evidence to the nonequivalence between the Hardy formula and the stress formula as well as to the nonequivalence between the LAMMPS implementation of the stress formula and our implementation based on the pairwise force. Particularly, we showed that the stress-based formulation underestimates the thermal conductivity of systems described by many-body potentials, and that this effect is more noticeable for low-dimensional systems. Our findings are very relevant for scientists  modelling thermal transport in low-dimensional systems via molecular dynamics simulations.

\begin{acknowledgments}
This research has been supported by the Academy of Finland through its Centres of Excellence Program (Project No. 251748). We acknowledge the computational resources provided by Aalto Science-IT project and Finland's IT Center for Science (CSC).  ZF acknowledges the support of the National Natural Science Foundation of China (Grant No. 11404033). LFCP acknowledges the provision of computational resources by the International Institute of Physics at UFRN. HQW and JCZ acknowledge the support of the National Natural Science Foundation of China (Grant No. U1232110), the Specialized Research Fund for the Doctoral Program of Higher Education (Grant No. 20120121110021), and the National High-tech R\&D Program of China (863 Program, No. 2014AA052202).
\end{acknowledgments}

\appendix

\section{\label{appendix:explicit_expression}Explicit expression for $\frac{\partial U_i}{\partial \vect{r}_{ij}}$}

In this appendix, we present an explicit expression for $\frac{\partial U_i}{\partial \vect{r}_{ij}}$, which can be easily implemented in a computer language.

Using the partition given by Eq.~(\ref{equation:U_i}), we have
\begin{equation}
\frac{\partial U_i}{\partial \vect{r}_{ij}}
= \frac{1}{2}\frac{\partial U_{ij}}{\partial \vect{r}_{ij}}
+ \frac{1}{2}\sum_{k\neq i,j}\frac{\partial U_{ik}}{\partial \vect{r}_{ij}}.
\end{equation}
After some algebra, we have
\begin{align}
\frac{\partial U_i}{\partial \vect{r}_{ij}}
&= \frac{1}{2}f_C'(r_{ij})[f_R(r_{ij})-b_{ij}f_A(r_{ij})]\frac{\partial r_{ij}}{\partial \vect{r}_{ij}} \nonumber \\
&+ \frac{1}{2}f_C(r_{ij})[f_R'(r_{ij})-b_{ij}f_A'(r_{ij})]\frac{\partial r_{ij}}{\partial \vect{r}_{ij}} \nonumber \\
&- \frac{1}{2}\sum_{k\neq i,j}f_C(r_{ik})f_C'(r_{ij})f_A(r_{ik})b'_{ik}g_{ijk}\frac{\partial r_{ij}}{\partial \vect{r}_{ij}} \nonumber \\
&- \frac{1}{2}\sum_{k\neq i,j}f_C(r_{ik})f_C(r_{ij})g'_{ijk}
   \frac{\partial \cos\theta_{ijk}}{\partial \vect{r}_{ij}} \nonumber \\
&\times [f_A(r_{ij})b'_{ij}+f_A(r_{ik})b'_{ik}],
\end{align}
where
\begin{equation}
\frac{\partial r_{ij}}{\partial \vect{r}_{ij}} = \frac{\vect{r}_{ij}}{r_{ij}},
\end{equation}
\begin{equation}
\frac{\partial \cos\theta_{ijk}}{\partial \vect{r}_{ij}}
= \frac{1}{r_{ij}}\left[\frac{\vect{r}_{ik}}{r_{ik}}
- \frac{\vect{r}_{ij}}{r_{ij}} \cos\theta_{ijk}\right],
\end{equation}
and we have used the following notations:
$f_A'(r_{ij}) \equiv \partial f_A(r_{ij}) / \partial r_{ij}$,
$f_R'(r_{ij}) \equiv \partial f_R(r_{ij}) / \partial r_{ij}$,
$f_C'(r_{ij}) \equiv \partial f_C(r_{ij}) / \partial r_{ij}$,
$b'_{ij} \equiv \partial b_{ij} / \partial \zeta_{ij}$, and
$g'_{ijk} \equiv \partial g_{ijk} / \partial \cos\theta_{ijk}$.

\section{\label{appendix:hardy}Unifying different heat current expressions in the literature}

The derivation of the heat current expressions for a general lattice has been considered very early by Hardy \cite{hardy1963} at the quantum level. The potential part of the heat current was derived to be
\begin{equation}
\label{equation:Hardy_quantum}
 \vect{J}_{\text{pot}}^{\text{Hardy}} = \frac{1}{2} \sum_i \sum_{j \neq i}
 \vect{r}_{ji} \frac{1}{i \hbar} \left[ \frac{\vect{p}_i^2}{2m_i}, U_j \right] + \text{h.c.},
\end{equation}
where $\hbar$ is reduced Planck constant, $\vect{p}_i$ and $m_i$ are the momentum operator and mass for particle $i$, and h.c. stands for Hermitian conjugate. Using the identity
\begin{equation}
 [\vect{p}_i, U_j] = -i\hbar \frac{\partial U_j}{\partial \vect{r}_i},
\end{equation}
the classical analog of Eq.~(\ref{equation:Hardy_quantum}) can be derived to be
\begin{equation}
\vect{J}^{\textmd{Hardy}}_{\textmd{pot}} = \sum_i \sum_{j \neq i} \vect{r}_{ij}
\left(
      \frac{\partial U_j}{\partial \vect{r}_{i}} \cdot \vect{v}_i
\right).
\end{equation}
Using Eq.~(\ref{equation:crucial_property_again}), we have
\begin{equation}
\frac{\partial U_{j}} {\partial \vect{r}_{i}}
= \sum_{k \neq j} \frac{\partial U_{j}} {\partial \vect{r}_{jk}}
\frac{\partial \vect{r}_{jk}} {\partial \vect{r}_{i}}
= \frac{\partial U_{j}} {\partial \vect{r}_{ji}},
\end{equation}
and
\begin{equation}
\vect{J}^{\textmd{Hardy}}_{\textmd{pot}} = \sum_i \sum_{j \neq i} \vect{r}_{ij}
\left(
      \frac{\partial U_j}{\partial \vect{r}_{ji}} \cdot \vect{v}_i
\right).
\label{equation:Hardy_classical}
\end{equation}
This equation is identical to Eq.~(\ref{equation:j_pot_tersoff_4}) and thus equivalent to all the expressions in Eqs.~(\ref{equation:j_pot_tersoff_1}-\ref{equation:j_pot_tersoff_3}).

We now show that many of the seemingly inequivalent expressions of the potential part of the heat current for the Tersoff/Brenner potential are equivalent to the Hardy formula.

We first consider the one used by Li \textit{et al.}, \cite{li1998} which takes the following form:
\begin{equation}
\vect{J}^{\textmd{Li}}_{\textmd{pot}} = - \sum_i \sum_{j \neq i}
          \vect{r}_{ij}
          \frac{\partial E_{i}} {\partial \vect{r}_{j}} \cdot \vect{v}_j.
\end{equation}
Since $\frac{\partial}{\partial \vect{r}_{j}}\left(\frac{1}{2}m_i\vect{v}_i^2\right)=\vect{0}$, we have
\begin{equation}
\label{equation:j_pot_tersoff_Li}
\vect{J}^{\textmd{Li}}_{\textmd{pot}} = - \sum_i \sum_{j \neq i}
          \vect{r}_{ij}
          \frac{\partial U_{i}} {\partial \vect{r}_{j}} \cdot \vect{v}_j,
\end{equation}
which has the same form as that used by Dong \textit{et al.}. \cite{dong2001} By noticing that [where we have used Eq.~(\ref{equation:crucial_property_again})]
\begin{equation}
\frac{\partial U_{i}} {\partial \vect{r}_{j}}
= \sum_{k \neq i} \frac{\partial U_{i}} {\partial \vect{r}_{ik}}
  \frac{\partial \vect{r}_{ik}} {\partial \vect{r}_{j}}
= \frac{\partial U_{i}} {\partial \vect{r}_{ij}},
\end{equation}
we have
\begin{equation}
\vect{J}^{\textmd{Li}}_{\textmd{pot}} = \vect{J}^{\textmd{Dong}}_{\textmd{pot}} =
        - \sum_i \sum_{j \neq i}
          \vect{r}_{ij}
          \frac{\partial U_{i}} {\partial \vect{r}_{ij}} \cdot \vect{v}_j.
\end{equation}
which is exactly Eq.~(\ref{equation:j_pot_tersoff_3}) and is thus equivalent to the Hardy formula. We also note that the one used by Berber \textit{et al.} \cite{berber2000} is exactly the Hardy formula.

We next consider the one derived by Che \textit{et al.}, \cite{che2000} which takes the following form:
\begin{equation}
\vect{J}^{\textmd{Che}}_{\textmd{pot}} = - \frac{1}{2} \sum_i \sum_j \sum_k \sum_l
          \vect{r}_{ik}
          \frac{\partial U_{kl}} {\partial \vect{r}_{ij}} \cdot \vect{v}_i.
\end{equation}
Since
\begin{equation}
\frac{\partial U_{kl}} {\partial \vect{r}_{ij}}
= \sum_m \frac{\partial U_{kl}} {\partial \vect{r}_{km}}
(\delta_{ki} \delta_{mj} - \delta_{kj} \delta_{mi}),
\end{equation}
we have
\begin{align}
\vect{J}^{\textmd{Che}}_{\textmd{pot}} &= \frac{1}{2} \sum_i \sum_j \sum_l
          \vect{r}_{ij}
          \frac{\partial U_{jl}} {\partial \vect{r}_{ji}} \cdot \vect{v}_i \nonumber \\
        &= \sum_i \sum_{j \neq i}
          \vect{r}_{ij}
          \frac{\partial U_{j}} {\partial \vect{r}_{ji}}
          \cdot \vect{v}_i,
\end{align}
which is exactly the Hardy formula.

The Hardy formula is also equivalent to a seemingly different one derived by Chen \textit{et al.}, \cite{chen2006} which reads (The original expression in Ref.~[\onlinecite{chen2006}] contains a typo, which has been noticed by Guajardo-Cu\'ellar \textit{et al.}. \cite{cuellar2010})
\begin{align}
\vect{J}^{\textmd{Chen}}_{\textmd{pot}} =
          &- \frac{1}{2}\sum_i \sum_{j \neq i}
          \vect{r}_{ij} \frac{\partial U_{ij}} {\partial \vect{r}_{j}} \cdot \vect{v}_j \nonumber \\
          &-\frac{1}{2}\sum_i \sum_{j \neq i} \sum_{k \neq i,j}
          \vect{r}_{ik} \frac{\partial U_{ij}} {\partial \vect{r}_{k}} \cdot \vect{v}_k.
\end{align}
By a change of indices ($k\leftrightarrow j$), the second term on the RHS of the above equation can be written as
\begin{equation}
-\frac{1}{2}\sum_i \sum_{j \neq i} \sum_{k \neq i,j}
\vect{r}_{ij} \frac{\partial U_{ik}} {\partial \vect{r}_{j}} \cdot \vect{v}_j,
\end{equation}
which, combining with the first term, gives [using Eq.~(\ref{equation:U_i})]
\begin{equation}
\vect{J}^{\textmd{Chen}}_{\textmd{pot}} = - \sum_i \sum_{j \neq i}
          \vect{r}_{ij} \frac{\partial U_{i}} {\partial \vect{r}_{j}} \cdot \vect{v}_j.
\end{equation}
It takes the same form of Eq.~(\ref{equation:j_pot_tersoff_Li}) and is thus equivalent to the Hardy formula.

Recently, Guajardo-Cu\'ellar \textit{et al.} \cite{cuellar2010} also derived an expression for the potential part of the heat current. They have used the equation
$m_i \frac{d \vect{v}_i}{dt} =  \sum_{j \neq i} \frac{\partial U_{ij}}{\partial \vect{r}_{ij}}$
in their derivation, which means that the force on particle $i$ was taken to be $\vect{F}_i= \sum_{j \neq i} \frac{\partial U_{ij}}{\partial \vect{r}_{ij}}$. This is only valid for two-body potentials, and as such it is not valid for the Tersoff potential. We thus do not expect that their expression is equivalent to the Hardy formula.

Last, we notice that Li \textit{et al.} \cite{li1998} also presented the potential part of the heat current as the sum of the following parts:
\begin{equation}
\label{equation:j_pot_tersoff_Li_1}
\vect{J}^{\textmd{Li1}}_{\textmd{pot}} =
- \frac{1}{4}\sum_i \sum_{j \neq i}
\left(
\vect{r}_{ij} \frac{\partial U_{ij}} {\partial \vect{r}_{j}} \cdot \vect{v}_j
+\sum_{k \neq i,j}
\vect{r}_{ik} \frac{\partial U_{ij}} {\partial \vect{r}_{k}} \cdot \vect{v}_k
\right)
\end{equation}
and
\begin{equation}
\label{equation:j_pot_tersoff_Li_2}
\vect{J}^{\textmd{Li2}}_{\textmd{pot}} =
- \frac{1}{4}\sum_i \sum_{j \neq i}
\left(
\vect{r}_{ji} \frac{\partial U_{ij}} {\partial \vect{r}_{i}} \cdot \vect{v}_i
+\sum_{k \neq i,j}
\vect{r}_{jk} \frac{\partial U_{ij}} {\partial \vect{r}_{k}} \cdot \vect{v}_k
\right).
\end{equation}
It can be shown that $\vect{J}^{\textmd{Li1}}_{\textmd{pot}}=\vect{J}^{\textmd{Hardy}}_{\textmd{pot}}/2$ and
$\vect{J}^{\textmd{Li2}}_{\textmd{pot}} \neq \vect{J}^{\textmd{Hardy}}_{\textmd{pot}}/2$ if one assumes the partition of potential energy given by Eq.~(\ref{equation:U_i}). However, they have in fact chosen a different decomposition:
\begin{equation}
 \label{equation:U_i_Li}
 U_i = \frac{1}{4}\sum_{j\neq i} \left( U_{ij} + U_{ji} \right).
\end{equation}
The calculated thermal conductivity is usually insensitive to the specific decomposition of the potential energy, as shown by Schelling \textit{et al.}\cite{schelling2002} for SW silicon.


\begin{thebibliography}{99}
\bibitem{note_convention}
Note that different papers may have used a different
convention and the reader should be aware of this
when comparing our equations with others.
\end{thebibliography}
\end{document}